\documentclass[journal=apchd5,manuscript=article]{achemso}

\usepackage[version=3]{mhchem} % Formula subscripts using \ce{}
\usepackage{amsmath}

\usepackage{xr-hyper}  % package for hyper refrances needed for referesing SI
\usepackage{hyperref}
\author{Tom T. C. Sistermans}
\affiliation{Photonics Research Group, Ghent University-imec, Ghent, Belgium}
\alsoaffiliation{Center for Nano- and Biophotonics, Ghent University, Ghent, Belgium}
\alsoaffiliation{Department of Applied Physics and Eindhoven Hendrik Casimir Institute, Eindhoven University of Technology, Eindhoven, The Netherlands}
\author{Rasmus H. Godiksen}
\affiliation{Department of Applied Physics and Eindhoven Hendrik Casimir Institute, Eindhoven University of Technology, Eindhoven, The Netherlands}
\author{Sara A. Elrafei}
\affiliation{Department of Applied Physics and Eindhoven Hendrik Casimir Institute, Eindhoven University of Technology, Eindhoven, The Netherlands}

\author{Alberto G. Curto}
\affiliation{Photonics Research Group, Ghent University-imec, Ghent, Belgium}
\alsoaffiliation{Center for Nano- and Biophotonics, Ghent University, Ghent, Belgium}
\alsoaffiliation{Department of Applied Physics and Eindhoven Hendrik Casimir Institute, Eindhoven University of Technology, Eindhoven, The Netherlands}

\email{Alberto.Curto@UGent.be}

\title{Fluctuation imaging of disorder\\ in monolayer semiconductors}

\abbreviations{}

\begin{document}

%\begin{tocentry}

%\centering\includegraphics[height=3.5cm]{TOC.pdf}
%Some journals require a graphical entry for the Table of Contents.
%This should be laid out ``print ready'' so that the sizing of the
%Inside the \texttt{tocentry} environment, the font used is Helvetica
%8\,pt, as required by \emph{Journal of the American Chemical
%The surrounding frame is 9\,cm by 3.5\,cm, which is the maximum
%permitted for  \emph{Journal of the American Chemical Society}
%graphical table of content entries. The box will not resize if the
%content is too big: instead it will overflow the edge of the box.
%
%This box and the associated title will always be printed on a
%separate page at the end of the document.
%
%\end{tocentry}


\begin{abstract}

Monolayer semiconductors hold great potential for nanoscale electronics, optoelectronics, and photonics. Excitons dominate their optical properties. As their electric fields extend outside the monolayer, they are sensitive to their surroundings. Thus, disorder can cause exciton instability, which is detrimental to device performance and critical for scalability and reproducibility. Here, we adapt a super-resolution fluorescence fluctuation microscopy technique to image localized exciton fluctuations in monolayer semiconductors, allowing us to identify unstable spots in an otherwise continuous monolayer with constant fluorescence. These spots correspond to interfacial disorder measured by atomic force microscopy. We examine how different material interfaces influence the fluctuations by comparing several substrates and provide additional insight into the disorder behind the fluctuations using hyperspectral imaging. We also assess the reduction of disorder upon thermal annealing, evidenced by a decrease in fluctuations. Our results show that fluorescence fluctuation imaging can detect disorder features similar to those of atomic force microscopy and hyperspectral imaging, while being faster and easier to implement. Therefore, it is a promising method for evaluating the quality of monolayer semiconductors, particularly when integrated with nanostructures and heterostructures found in nano-optoelectronic devices.

\end{abstract}


\section*{Introduction}
% general introduction to TMDs

Transition metal dichalcogenides (TMDs) are layered van der Waals materials that can be thinned down to semiconducting monolayers by exfoliation. TMDs are promising for nanoscale electronics, optoelectronics, and photonics due to their unique properties arising from their two-dimensional (2D) nature and carrier dynamics~\cite{Das_2021,perspective2025}. Excitons, which are bound electron-hole pairs, exhibit increased binding energy due to their 2D character (e.g., 0.71 eV for WS$_2$~\cite{Zhu2015,Splendiani2010}), leading to a dominant role in the optical and electronic properties even at room temperature. The exciton wavefunction extends beyond the monolayer, allowing for a strong interaction with the surrounding environment~\cite{Rahman2021} and making monolayer TMDs also suitable for sensing applications with electronic or optical readout~\cite{Hu2017, Reynolds2019}. However, this sensitivity also implies that the quality of the monolayers and their interfaces critically influences their properties within devices~\cite{Chen2021}. Therefore, industrial scaling of 2D semiconductor electronics and photonics requires the development of metrology techniques for quality assessment~\cite{perspective2025}.

% introducing disorder
In practice, TMD monolayers and their interfaces are not perfect, as disorder can manifest in various forms, such as substrate roughness, wrinkles, impurities, and crystal lattice defects. Such disorder affects the material properties, including band structure, exciton binding energy, emission linewidth, and quantum efficiency~\cite{Raja2019,Raja2017,Darlington2020,Harats2020,Reynolds2019,Lin2016}. Different techniques are available to study disorder in 2D semiconductors. Hyperspectral imaging has been used to map the exciton emission spectrum, providing insights into local excitonic properties~\cite{Wang2022}. The effect of strain caused by nanobubbles was studied with near-field imaging, providing high-resolution maps of photon energy alongside atomic force microscopy (AFM) height~\cite{Darlington2020}. Crystal defects, such as chalcogen vacancies, have been studied by transmission electron microscopy, scanning tunneling microscopy, and fluorescent labeling, revealing the presence of grain boundaries and a decrease in quantum efficiency at high defect densities~\cite{Lin2015,Vancso2016,Zhang2021}. 

% Introducing fluctuations and show the current state of the art
Besides spatial variations caused by disorder, TMD emission can also exhibit temporal fluctuations. Previously, spatially extended fluorescence blinking of entire monolayer TMD domains has been reported, comparable to the blinking of 0D quantum emitters but occurring in 2D~\cite{Godiksen2020,ye2017,Ai2011,Zhou_2025}. Blinking over extended areas has been reported in TMD monolayer heterostructures~\cite{huangMoireBlinking2025, HuaInterfaces2023}, where both monolayers can fluctuate in anticorrelation~\cite{Xu2017}. Like conventional quantum dots, quantum emitters in monolayer TMDs also show intermittent localized fluorescence at low temperatures~\cite{He2015}. At room temperature, TMDs were shown to display blinking or flickering due to energy transfer between single quantum dots and MoS$_2$~\cite{He2022}, as well as between perovskite nanowires and MoSe$_2$~\cite{LuoMoSe2Perovskite2024}. Other 2D materials like hexagonal boron nitride (hBN) display fluorescence activated by the environment, resulting in quantum emitters with a spectral distribution that contains information on the permittivity of the surrounding liquid~\cite{Ronceray2023}, and enabling monitoring of electrochemical dynamics~\cite{Mayner2024}. Similarly, molecular fluorescence has been used to detect electronic puddles in graphene~\cite{ciancico2025opticaldetectionchargedefects}.
In light of these developments, methods for imaging, quantifying, and leveraging the information contained in fluorescence fluctuations in monolayer materials are needed. 

% methods of fluctuation sensing

In super-resolution microscopy, fluctuations are commonly used to extract additional information about the imaged system, enabling resolution beyond the diffraction limit. One method that relies on fluctuations for imaging is super-resolved optical fluctuation imaging (SOFI), which produces an image where the initial brightness is weighted by the fluctuation strength~\cite{Dertinger2009}. SOFI thus results in images with a strongly reduced background, while blinking emitters are enhanced, accompanied by a modest increase in resolution. Other methods for super-resolution imaging, like stochastic optical reconstruction microscopy (STORM), photoactivated localization microscopy (PALM), and point accumulation for imaging in nanoscale topography (PAINT), rely on fluorescence switching or blinking to localize single emitters and can reach a significantly higher resolution than SOFI~\cite{Alva2022,Petrasek2009,Yahiatene2015}. However, they impose strong constraints on the emitter density and switching characteristics, rendering these methods impractical for label-free imaging of TMD monolayers, where spatially extended fluorescence acts as background for the fluctuations. 

% SUMMARY
Here, we exploit fluctuation imaging using SOFI to map and quantify localized fluorescence fluctuations in monolayer WS$_2$. The resulting fluctuation strength map identifies disorder sites in the monolayer that correlate with the disorder found in AFM images.
We further investigate the origin of the fluctuating spots using hyperspectral imaging and compare fluctuation images for different substrates. We also observe a reduction in fluctuations after thermal annealing. The resulting disorder maps reveal detailed information about the semiconductor and its interfaces, 
establishing fluctuation imaging as a sensitive indicator of monolayer material quality with applications in semiconductor metrology and sensing.

\section*{Super-resolved optical fluctuation imaging}

%short introduction

For an extended 2D emitter such as a monolayer imaged on a camera, we can write the fluorescence detected by a pixel as a function of time as $F(t)=\epsilon s(t)$, where $\epsilon$ is the time-averaged intensity $\langle F(t) \rangle_t$ and $s(t)$ is a time-dependent fluctuation. It can be expressed as a zero-mean fluctuation using $\delta F(t)=F(t)-\langle F(t)\rangle_t=\epsilon(s(t)-1)$. The autocorrelation function as a function of time delay, $G(\tau)$, computed for second-order SOFI is then given by \cite{Dertinger2009}:
\begin{equation}
    G(\tau)=\epsilon^2 \langle (s(t+\tau)-1)(s(t)-1)\rangle_t.
\end{equation}
This correlation function depends on the intensity squared, which reduces the contrast for dim emitters compared to bright ones. To linearize this result in intensity, we take the square root of the correlation function. The resulting linearized SOFI image effectively shows the initial intensity, weighted by a term that depends on the fluctuation strength. Additionally, similarly to autocorrelations, cross-correlations between neighboring pixels are used to compute virtual pixels between real pixels~\cite{Geissbuehler2011}. 

To decouple the fluctuation strength from fluorescence intensity, we introduce a variant of SOFI tailored for fluctuation quantification imaging, which we call quantitative SOFI (qSOFI). We define the qSOFI signal as:
\begin{equation}
    qSOFI(\tau) = \frac{SOFI_\mathrm{lin}(\tau)}{\langle F(t) \rangle_t}
    =  \sqrt{\langle (s(t+\tau)-1)(s(t)-1)\rangle_t} ,
\end{equation}
which depends only on $s(t)$ and its delayed version, and is thus independent of mean intensity (see Supporting Section~\ref{SI-sec:math_qSOFI} for details). In contrast to conventional SOFI, the effective spatial resolution of qSOFI depends on the relative contribution of the fluctuation and intensity terms. Features dominated by localized fluctuations inherit the super-resolved characteristics of the SOFI image, whereas features primarily originating from spatial variations in the underlying PL intensity remain limited by the diffraction-limited PL map. Examples illustrating both cases are provided in Supporting Section~\ref{SI-sec:resolution}. In this work, we employ second-order qSOFI because our primary objective is to identify fluctuation sites by suppressing the nearly constant PL background rather than to achieve the highest possible spatial resolution. Although higher-order SOFI can further increase resolution, it generally requires larger datasets and higher signal-to-noise ratios. Second-order qSOFI therefore provides a suitable balance between fluctuation sensitivity and image robustness for the present study.

The time delay, $\tau$, is a parameter that allows SOFI to be tuned to different fluctuation time scales. Increasing the time delay in qSOFI removes fast fluctuations, which are incoherent at longer delays and will thus average out for sufficiently large datasets. In general, the time delay is a parameter that can be optimized for specific cases. For small datasets, it can prove beneficial to integrate over a range of time delays. 

In the limit of zero time delay, qSOFI is related to the standard deviation normalized by the mean intensity, as both quantify the relative strength of fluorescence fluctuations. A direct comparison between qSOFI and standard-deviation-based fluctuation imaging is provided in Supporting Section~\ref{SI-sec:std_comparison}. By combining auto-correlations and cross-correlations between neighboring pixels, SOFI generates virtual pixels and provides a spatial-resolution enhancement, improving the separation and visibility of nearby fluctuation sites. Furthermore, the cross-correlation-based virtual pixels are insensitive to uncorrelated shot noise, which otherwise contributes background to fluctuation measurements. In addition, SOFI processing can incorporate drift and bleaching correction~\cite{sofipackage} and can be extended to higher-order correlations for further resolution enhancement~\cite{Dertinger2009}. As a result, qSOFI retains the intuitive interpretation of a normalized fluctuation map while benefiting from the additional spatial resolution and processing features available within the SOFI framework.

\begin{figure}[tb]
  \centering\includegraphics[width=0.85\textwidth]{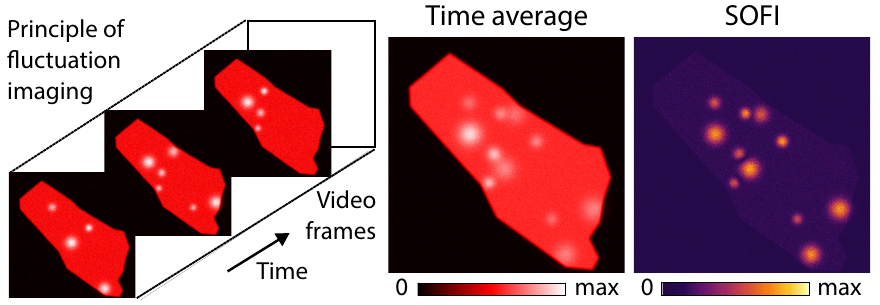}
  \caption{\textbf{Optical fluctuation imaging for monolayer semiconductors.} Schematic video frames of a two-dimensional emitter with fluctuating bright spots on a constant background. Average fluorescence intensity and linearized SOFI images calculated from the illustrative frames, which show a strong suppression of background fluorescence and an increased visibility of the fluctuating spots.
  \label{fig1_SOFI_scematic}}
\end{figure}

The fluorescence of monolayer semiconductors consists of a time-averaged background plus localized, fluctuating spots, as we will demonstrate experimentally later. We illustrate imaging of such an object using SOFI in Figure~\ref{fig1_SOFI_scematic}. Starting with schematic video frames of an area with constant fluorescence containing localized fluctuating spots, we process the image stack using both time averaging and linearized SOFI. The SOFI image shows the ability to locate fluctuating spots while suppressing the near-constant background.

\section*{Imaging disorder in semiconductor monolayers}

\begin{figure}[tb]
  \centering\includegraphics[width=0.9\textwidth]{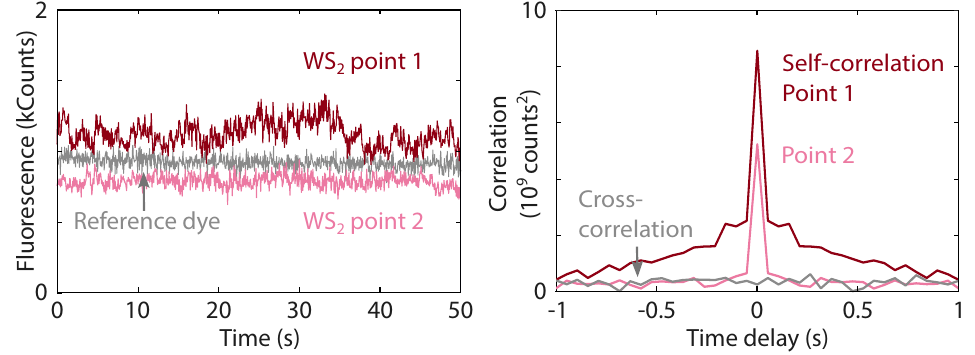}
  \caption{\textbf{Localized and independent fluorescence fluctuations for two points on a WS$_2$ monolayer.} Left: measured time traces showing distinct fluorescence fluctuations for two points on the same monolayer on an ITO substrate, separated by 4~$\mu$m, compared to a stable fluorescent dye. Right: autocorrelation of the time traces of points 1 and 2 and cross-correlation between them as a function of time delay, proving that the localized fluctuations are independent despite being on the same monolayer. Both points are located on the sample in Figure~\ref{fig3_intro_qSOFI}.
  \label{fig2_time_trace}}
\end{figure}

\begin{figure}[tb]
  \centering\includegraphics[width=0.85 \textwidth]{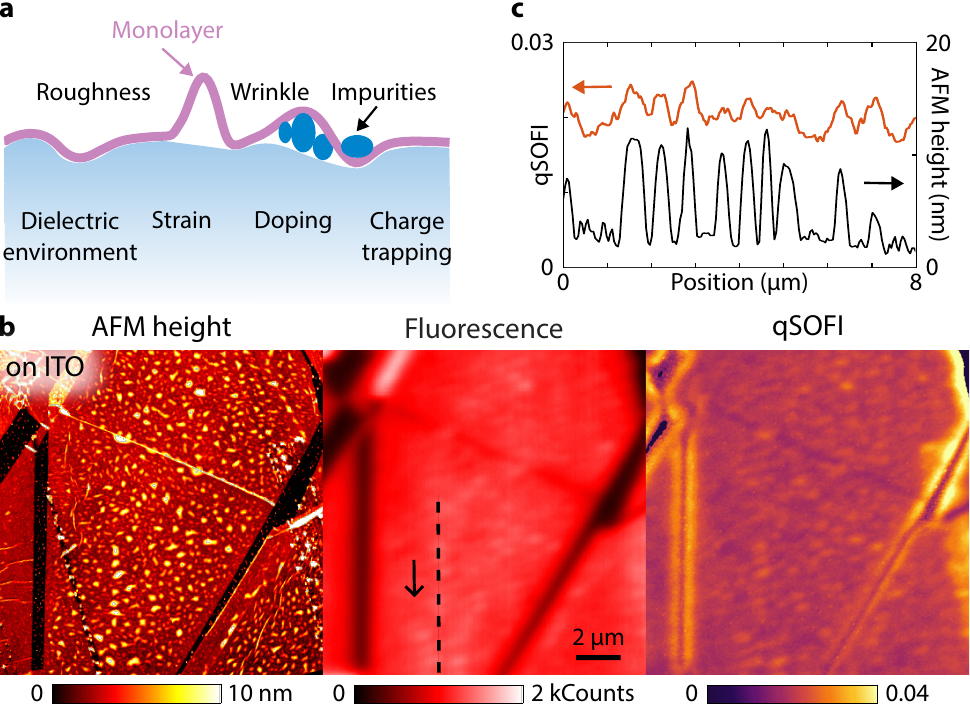}
  \caption{\textbf{Fluctuation imaging reveals disorder in a WS$_2$ monolayer.} (a) Sources of disorder in a monolayer and their effects. (b) Comparison between height, time-averaged fluorescence, and qSOFI for a monolayer WS$_2$ on an ITO-coated substrate. (c) Cross-section along the dashed line in (b) comparing the height profile and the qSOFI value. \label{fig3_intro_qSOFI}}
\end{figure}

In our experiments, we image fluctuations in a wide-field fluorescence microscope equipped with a blue lamp and a sCMOS camera (see Methods). The acquired videos are post-processed into fluctuation images using qSOFI, which allows us to experimentally identify fluctuating spots on a WS$_2$ monolayer (points 1 and 2 in Figure~\ref{fig2_time_trace}, part of the monolayer displayed in Figure~\ref{fig3_intro_qSOFI}). The fluorescence time traces at these two points show that point 1 is brighter and fluctuates more strongly than point 2 (Figure~\ref{fig2_time_trace}). Note that both points exhibit fluctuations above the instrument noise level measured with a reference sample containing stable fluorescent dye molecules (Supporting Section~\ref{SI-sec:stable_ref}).
% New Pink BASF sample
The fluctuation strength determined by SOFI is based on the temporal correlation in Figure \ref{fig2_time_trace}. The correlation at zero time delay is higher and decays more slowly for point 1 than for point 2, indicating stronger and slower fluctuations. Both points fluctuate independently, as demonstrated by the lack of cross-correlation at all times, which reflects their localized nature.

TMD monolayers are often depicted as a perfect 2D crystal lying flat on a smooth substrate. In practice, however, the situation is usually far more complex with a variety of causes for disorder. The exfoliation process typically used to obtain monolayers can introduce wrinkles and trap air bubbles between the monolayer and the substrate. Interfaces are critical for 2D materials, and both substrate roughness and residue introduced during sample preparation locally alter monolayer properties~\cite{Raja2019}. Crystal defects, particularly chalcogen vacancies, likewise affect the monolayer optical response~\cite{Zhang2021,Zehua2018}. The inhomogeneities introduced by fabrication can thus result in spatially varying dielectric environments, strain, doping, and charge-trapping densities, as schematically represented in Figure \ref{fig3_intro_qSOFI}a. These variations in a TMD monolayer modulate the fluorescence intensity, and hence the excitonic emission reflects the inhomogeneous landscape as localized temporal fluctuations.

AFM provides high-resolution imaging of disorder, as illustrated in Figure \ref{fig3_intro_qSOFI}b for a WS$_2$ monolayer on indium tin oxide (ITO). The fluorescence intensity image of the same region hints at the underlying disorder, but the combination of background fluorescence from the continuous monolayer and the diffraction-limited resolution blurs the disorder features. Using qSOFI, we process the corresponding fluorescence video to compute a fluctuation image. An analysis of the reconstruction variability and signal-to-noise ratio is provided in Supporting Section~\ref{SI-sec:qSOFI_noise}. The resulting qSOFI image makes individual disorder spots distinguishable by increasing contrast through background suppression. Comparing line profiles across the monolayer in the qSOFI and AFM images reveals a high correlation between regions with roughness and elevated fluctuation strength (Figure \ref{fig3_intro_qSOFI}c). Over a larger area within this monolayer, the Pearson correlation coefficient is 0.55, demonstrating a correlation between similar features in both images. In contrast, conventional fluorescence imaging yields a correlation coefficient of only 0.18, indicating that qSOFI more effectively highlights disorder-related features correlated with the AFM morphology. As AFM and qSOFI measure different quantities and the physical mechanisms leading to fluctuations are diverse, a height variation does not necessarily map directly onto fluctuation strength, even though both images show remarkable similarities. Both the electronic and morphological properties of the disordered monolayer environment influence exciton dynamics and, therefore, the fluctuation strength, requiring additional examination to be fully understood. This complexity was recently underscored by the reporting of different fluctuation statistics~\cite{Zhou_2025}.

To shed light on the origin of the localized fluctuations, we focus next on the analysis of spatially resolved spectral information. Neutral and charged excitons (trions) dominate the emission spectrum, and their relative peak positions and intensities can reveal further insight into the nanoscale origin of the disorder~\cite{bonnet2021}. Disorder can stem from different causes: strain resulting from bubbles and wrinkles; different dielectric environments caused by residue or large bubbles; doping due to residue or charge puddles; charge trapping at defect sites; photo-induced effects; substrate charging; and water or oxygen absorption on the monolayer.
Tensile strain decreases the bandgap while having little effect on the exciton binding energy, resulting in an overall red shift of exciton emission~\cite{Aslan2018}. The dielectric environment has a smaller impact on the A exciton because changes in the bandgap are partially offset by opposing changes in exciton binding energy~\cite{Yang2019,Hsu2019,Raja2019}. Disorder in both the dielectric environment and strain locally shifts the exciton emission energy, resulting in an increased inhomogeneous linewidth in the emission spectrum~\cite{Tweedie2018,Moody2015}. Although trions lie at a lower energy than neutral excitons, the emission is dominated by neutral excitons because the trion binding energy is comparable to the thermal energy at room temperature. However, free electrons are efficiently funneled by the strong band bending caused by disorder in strain and dielectric environment, leading to locally varying exciton-trion conversion rates~\cite{Harats2020,Lee2022}. The formation of trions is also affected by doping caused by residual molecules on the substrate and monolayer. Additionally, our WS$_2$ crystals are unintentionally $n$-type doped. The doping level affects the emission quantum efficiency because neutral excitons are converted to trions, which have a fast non-radiative component~\cite{Rahaman2021,Zhou2013,Wang2019}. Lastly, exciton trap states commonly associated with crystal defects and interfaces directly modulate fluorescence fluctuations~\cite{Zhu2023,Godiksen2020}.

\begin{figure}[tb]
  \centering\includegraphics[width=0.85\textwidth]{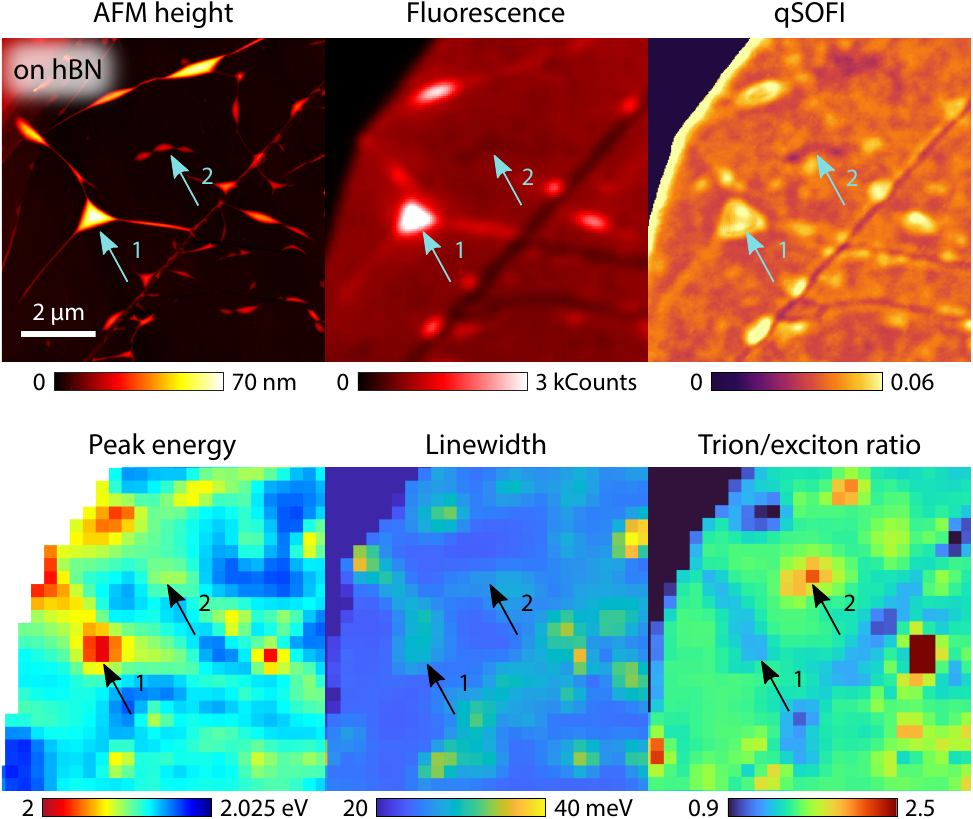}
  \caption{\textbf{Mapping disorder in a WS$_2$ monolayer on hBN through changes in exciton properties.} Top: height measured by AFM, time-averaged fluorescence, and fluorescence fluctuations shown as qSOFI. Bottom: exciton emission peak energy, linewidth, and trion intensity relative to neutral excitons retrieved from hyperspectral images of the same areas. Arrows point to two disordered areas showing opposing trends in relative fluorescence intensity, fluctuation strengths, and trion-to-neutral-exciton intensity ratios. \label{fig4_hyperspectral}}
  
\end{figure}

To illustrate the coexistence of these diverse disorder mechanisms, we employ hyperspectral imaging of a monolayer on another commonly used dielectric: hexagonal boron nitride (hBN), known for its potential for high-quality interfaces with monolayer semiconductors. Our transfer method resulted in a lower overall spot density on hBN (Figure~\ref{fig4_hyperspectral}) while introducing a few larger defects~\cite{Wierzbowski2017}, enabling us to characterize individual disorder features with diffraction-limited hyperspectral confocal imaging. We note that similar fluctuation images can be obtained for WSe$_2$ on hBN (Supporting Section~\ref{SI-sec:wse2}). In addition, measurements on multiple WS$_2$ monolayer flakes on hBN show qualitatively similar fluctuation patterns, demonstrating that the observed behavior is reproducible across different flakes (Supporting Section~\ref{SI-sec:extra_flakes}). Using spectral fitting, we extract the emission peak energy, linewidth, and trion-to-neutral-exciton ratio (Supporting Section~\ref{SI-sec:sup_spectral_fitting}).
The exciton emission peak energy shows a red shift at spots identified by AFM (compare top and bottom rows in Figure~\ref{fig4_hyperspectral}), which is consistent with the presence of local strain. Variations in peak energy are also accompanied by an increase in linewidth, suggesting inhomogeneous broadening in these regions. Furthermore, the trion-to-exciton intensity ratio varies significantly at disordered spots, which may be related to local variations in exciton-to-trion conversion, for instance through electron funneling~\cite{Harats2020,Lee2022}. Notably, some areas show an increase in the trion-to-exciton ratio while others show a decrease, possibly reflecting competing contributions from changes in the dielectric environment and strain, both of which can be present in large bubbles and wrinkles~\cite{Wang2022}.

Combining all the images in Figure~\ref{fig4_hyperspectral} demonstrates the presence of different types of disorder affecting emission.
These diverse physical mechanisms, often with counteracting effects, complicate pinpointing the origin of disorder at individual spots. As a result, qSOFI should be viewed as a method for identifying and quantifying disorder-related fluctuations rather than as a standalone tool for assigning a unique microscopic origin to a given fluctuation site.
Complementary characterization methods such as Kelvin probe force microscopy (KPFM)~\cite{Dappe2020}, near-field scanning microscopy~\cite{Darlington2020}, lifetime imaging~\cite{Barker2019}, and temperature-dependent measurements of the Stokes shift~\cite{Masenda2021} could provide such insight. Regardless of their origin, fluctuation imaging provides a method to identify disordered areas and compare the quality of monolayers before and after processing, as we demonstrate next.

\section*{Fluctuations for characterizing monolayer quality}
\begin{figure}[tb]
  \centering\includegraphics[width=0.85\textwidth]{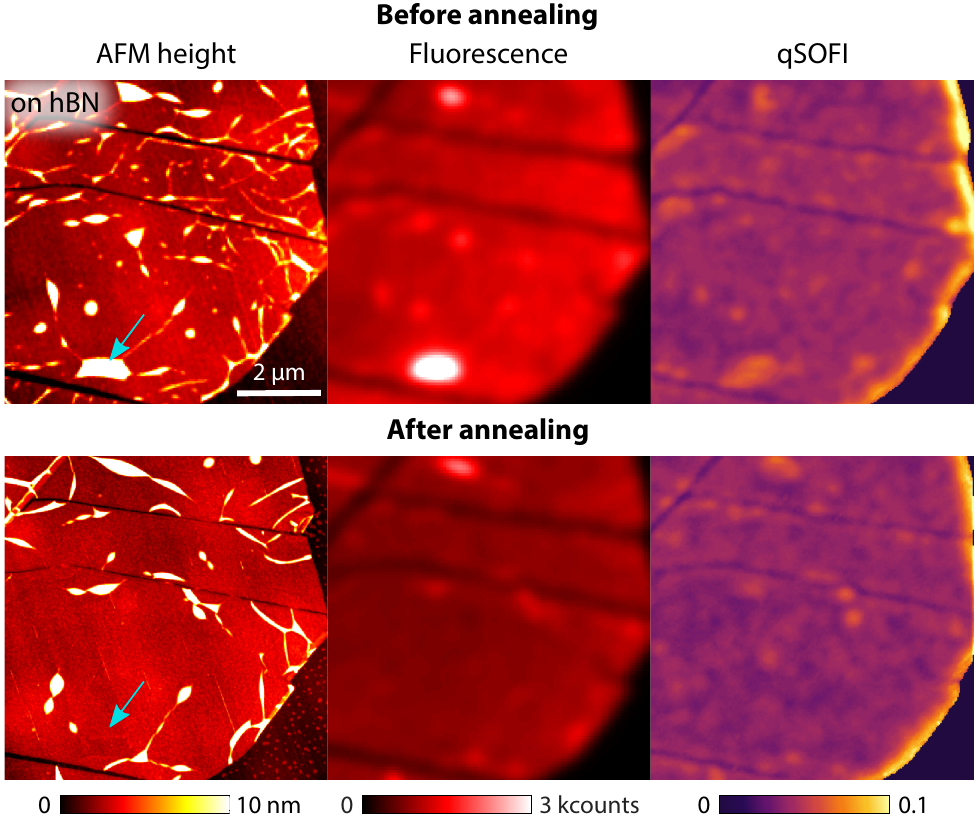}
  \caption{\textbf{Effect of annealing on disorder for a WS$_2$ monolayer on an hBN substrate.} Top and bottom rows: same monolayer before and after annealing. Small wrinkles disappear while some bubbles combine to form larger ones.\label{fig5_annealing_on_hBN}}
\end{figure}

Mitigating disorder introduced during fabrication is of practical importance for electronic, optoelectronic, and photonic devices ~\cite{Chen2021}.
Such disorder can be reduced by methods like chemical cleaning and thermal annealing~\cite{Carmiggelt2020}.
One major cause of disorder is the formation of bubbles and wrinkles, for example during the transfer of monolayers onto hBN using PDMS stamping. Thermal annealing is a promising strategy to mitigate these structural inhomogeneities. We apply fluctuation imaging to visualize the changes in quality of a WS$_2$ monolayer on hBN before and after thermal annealing under vacuum at 120\textdegree C (see Methods). Annealing allows bubbles and wrinkles to move and merge below the monolayer, where many of the smaller wrinkles disappear (AFM maps in Figure~\ref{fig5_annealing_on_hBN} and Supporting Section~\ref{SI-sec:annealing}). Fluorescence becomes more spatially homogeneous but decreases in overall intensity, consistent with accelerated aging induced by annealing~\cite{Gao2016}. The increase in homogeneity is also evident in fluctuation imaging, with fewer and more uniform features. Furthermore, the qSOFI value is slightly reduced, indicating weaker fluctuations and improved interface quality.

\begin{figure}[tb]
  \centering\includegraphics[width=0.85\textwidth]{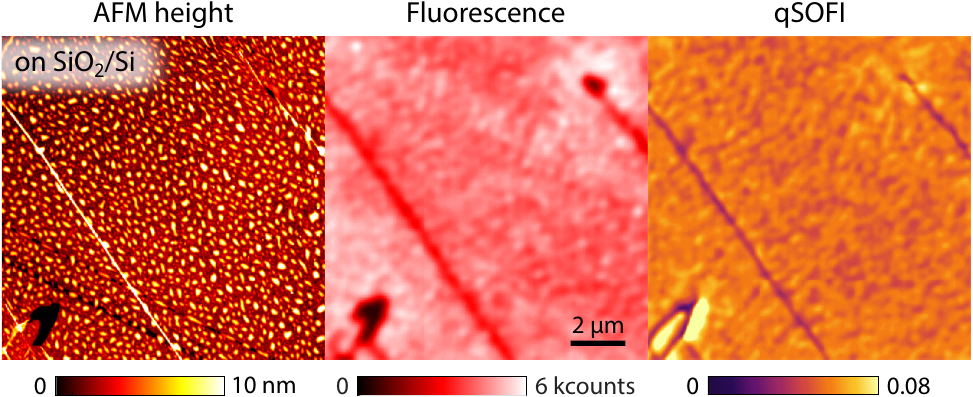}
  \caption{\textbf{Disorder in a WS$_2$ monolayer on an oxidized Si substrate.} From left to right: maps of the height profile measured by AFM, time-averaged fluorescence, and qSOFI.\label{fig6_ox_Si}}
  
\end{figure}

Finally, we employ our fluctuation imaging technique with two other commonly used substrate materials: oxidized silicon (SiO$_2$/Si) and PDMS.
The choice of substrate influences both the quality of the interfaces and, in some cases, also the monolayer transfer method. As we have already seen, using ITO substrates led to small disorder spots in Figure~\ref{fig3_intro_qSOFI}, whereas using hBN required an extra stamping step, resulting in bubbles and wrinkles in Figures ~\ref{fig4_hyperspectral} and ~\ref{fig5_annealing_on_hBN}. Using an oxidized silicon (SiO$_2$/Si) substrate, with the same exfoliation and transfer process as for ITO, yields similar fluctuation images in Figure~\ref{fig6_ox_Si}. However, the disorder spots are smaller and more densely distributed. This high density renders individual spots difficult to resolve with conventional diffraction-limited imaging, leading to spatial blurring and a loss of disorder localization in fluorescence intensity maps, whereas fluctuation imaging improves the visibility of some features. Similar fluctuation patterns are observed across multiple monolayer flakes on oxidized silicon (Supporting Section~\ref{SI-sec:extra_flakes}), demonstrating the reproducibility of these observations. Lastly, we investigate disorder on PDMS, which is sometimes used during exfoliation and stamping but can also serve as a substrate, eliminating a step in sample preparation. PDMS yields a relatively homogeneous monolayer also in fluctuation images (Supporting Section~\ref{SI-sec:pdms}), in agreement with its excellent suitability as a substrate for WS$_2$ monolayers, enabling narrow exciton linewidths and high oscillator strengths~\cite{Elrafey2024}. As the cases of PDMS and SiO$_2$/Si exemplify, fluctuation imaging is also a promising technique for identifying TMD-substrate combinations and fabrication methods that reduce disorder for exciton-based sensing applications with optimal performance. While the present study focuses on mechanically exfoliated monolayers, the approach could in principle be extended to CVD-grown TMDs and van der Waals heterostructures. The higher defect density and presence of grain boundaries in such materials may introduce stronger fluctuations.

\section*{Conclusions}

We have demonstrated that the fluorescence of TMD monolayers exhibits localized fluctuations above the noise level. Unlike the spatially extended, correlated blinking reported in previous works, these fluctuations are confined to nanoscale regions and thus provide information about the immediate interfacial environment of the monolayer. Disorder in the form of spatially varying dielectric permittivity, strain, doping, or defects affects the fluorescence intensity and its fluctuations.
We have proved that SOFI, a super-resolution microscopy technique based on fluctuations, can enable high-resolution mapping of monolayer semiconductors in realistic, imperfect environments. By adapting SOFI to quantify fluctuation strength independently of fluorescence intensity, we have shown that qSOFI highlights fine details of disorder in monolayers. 

Fluctuation imaging can localize disorder in light-emitting monolayer materials with a higher signal-to-background ratio than time-averaged fluorescence imaging. We compared fluctuation imaging with AFM and established a correlation between morphological disorder and fluctuation strength.
We investigated the physical origin of the disorder using hyperspectral imaging to probe local excitonic properties, demonstrating the coexistence of different sources of disorder leading to changes in peak emission energy, linewidth broadening, and trion-to-neutral-exciton ratio, often with competing effects. As application cases, we investigated the impact on fluctuation maps of thermal annealing for interface cleaning and the use of different material combinations, particularly substrates. 

Monolayer semiconductors form the basis of a wide range of technologies that rely on reproducible, high-quality monolayers, making metrology methods tailored to 2D semiconductors an essential requirement for their industrial scalability in electronics and photonics~\cite{perspective2025}.
Fluctuation imaging could function as an easy-to-implement method for quality control during the manufacturing process, requiring only a conventional wide-field fluorescence microscope and offering faster characterization than scanning probe techniques such as AFM. The method can also be more broadly applied to other monolayer semiconductors and excitonic nanomaterials beyond TMDs. Finally, since fluctuations report useful information about the local environment, quantifying them is also of interest for nanoscale sensing applications. 
The sensitivity of extended monolayer semiconductors to their surroundings has already been used in sensing applications~\cite{Hu2017,Reynolds2019}. Our results suggest that localized fluctuations could exhibit an even stronger dependence of fluorescence for sensing applications, enabling higher spatial resolution for mapping nanoscale analyte distributions.

\section*{Methods}

\subsection*{Sample preparation}

All WS$_2$ monolayers were exfoliated with Nitto blue tape (SPV224) from a bulk crystal (HQ Graphene) onto PDMS (Gel-Pak PF-80-X4). For the preparation of the samples on ITO (100 nm ITO on glass, Ossila S111) and SiO$_2$/Si (285 nm thermal oxide SiO$_2$ on Si, N-type, P-doped, MTI Corp.), the substrates were cleaned with acetone, methanol, and isopropanol under sonication for 5 minutes in each solvent. Then, the PDMS slab was placed on the substrate and left in a vacuum desiccator for more than 24 hours. The substrate/monolayer/PDMS stack was then heated to 70\textdegree C on a hot plate for 10 minutes and left to cool in air, allowing the PDMS to expand and contract, releasing the monolayers. Finally, the PDMS was slowly peeled off. The sample preparation of WS$_2$ on hBN started with exfoliation and transfer of hBN to SiO$_2$/Si, prepared in the same way as the WS$_2$ on SiO$_2$/Si samples. The hBN sample was then annealed for 2 hours at 120\textdegree C in a vacuum oven. The exfoliated WS$_2$ on PDMS sample was then stamped onto the hBN flakes under a microscope~\cite{Barker2019}, resulting in a Si/SiO$_2$/hBN/monolayer/PDMS stack. The PDMS was removed after heating. An example of this stack is shown in Supporting Section~\ref{SI-sec:stamping}. For the thermal treatment tests, WS$_2$ monolayers were annealed in a vacuum oven for 2 hours at 120\textdegree C. Subsequent measurements were performed under ambient conditions, and samples were stored in a vacuum desiccator.

\subsection*{Optical fluctuation imaging and processing}

We recorded fluorescence videos using a wide-field microscope with a blue lamp (Nikon Intensilight), a set of longpass dichroic mirror and filter, a microscope objective (Nikon TU Plan Fluor $100\times$, 0.9 NA), a tube lens, and an sCMOS camera (Andor Zyla 4.2 Plus). The camera captured 5000 frames with 50 ms integration time and background subtraction. For SOFI processing, we use the "SOFI Package", a MATLAB-based open-source project~\cite{sofipackage}. Processing was done on 500-frame subsequences to prevent long-term effects, such as defocusing and bleaching, from competing with faster but small fluorescence fluctuations. Drift correction between individual frames was used to compensate for microscope vibrations. To test this method, we imaged two closely spaced QDs, as shown in Supporting Section~\ref{SI-sec:resolve_QD}. For monolayers, the qSOFI images were obtained by dividing the second-order linearized SOFI by the time-averaged image with interpolated virtual pixels~\cite{Dertinger2009}. Note that qSOFI values at monolayer boundaries can be affected by normalization to a low signal. To exclude background regions outside the monolayer, a mask was generated from the time-averaged PL intensity image using a threshold of 20\% of the mean intensity. This threshold was used only to define the valid analysis area for visualization purposes. The qSOFI values themselves were calculated independently of the threshold and were not modified within the masked region.

\subsection*{Hyperspectral imaging}
Hyperspectral imaging was performed on a custom sample-scanning confocal microscope, using the same objective as for fluctuation imaging, under excitation with a 532-nm laser (Cobolt Samba) with laser power in the 1-10 $\mu$W range at the sample plane. The emitted light was collected by a fiber with a 50-$\mu$m core diameter and guided to a spectrometer (Andor Shamrock 303i) with a diffraction grating with 300 lines per mm and an EMCCD camera (Andor Newton 970). The obtained spectra were then fitted using a model with two Gaussians. An example of this fitting is shown in Supporting Section~\ref{SI-sec:sup_spectral_fitting}.

\subsection*{Atomic force microscopy}
AFM images were obtained with a Bruker Dimension Icon-PT Scanning Probe Microscope using ScanAsyst cantilevers. The AFM data was processed using flattening and zero $z$ offset in Gwyddion.

\begin{acknowledgement}
This work was financially supported by the European Research Council (ERC) under the European Union's Horizon 2020 Research and Innovation Program (Grant Agreement 948804, CHANSON).

\end{acknowledgement}

\begin{suppinfo}

Mathematical description of qSOFI; effective spatial resolution of qSOFI; comparison with standard-deviation-based fluctuation imaging; qSOFI imaging of a stable fluorescent reference; noise analysis and signal-to-noise ratio of qSOFI imaging; Imaging fluctuations in a tungsten diselenide monolayer; fluctuation imaging of additional flakes on the same substrate; emission spectral fitting; vacuum annealing to reduce disorder; fluctuation imaging using a PDMS substrate; stamping of monolayers on hBN; and super-resolved optical fluctuation imaging of two closely spaced quantum dots.

\end{suppinfo}

\bibliography{references}
%\newpage
%\listoftodos[Notes]
\end{document}